\documentstyle[twocolumn,prl,epsf,floats,aps]{revtex}
\newcommand{\mean}[1]{\langle #1 \rangle}
\newcommand{\ad}{a^{\dagger}}
\newcommand{\bd}{b^{\dagger}}
\newcommand{\fd}{f^{\dagger}}
\newcommand{\gd}{g^{\dagger}}
\newcommand{\be}{\begin{equation}}
\newcommand{\ee}{\end{equation}}
\newcommand{\bea}{\begin{eqnarray}}
\newcommand{\eea}{\end{eqnarray}}
\begin{document}
\draft
\hyphenation{Fok-ker}
\hyphenation{Chal-ker}
\hyphenation{Doh-men}
\preprint{}
\title{Scaling and Crossover Functions for the Conductance in the Directed
Network Model of Edge States}
\author{Ilya A. Gruzberg, N. Read and Subir Sachdev}
\address{Department of Physics, Yale University \\
P.O. Box 208120, New Haven, CT 06520-8120, USA.\\}
\date{November 22, 1996}
\maketitle

\vskip -5mm

\begin{abstract}

We consider the directed network (DN) of edge states on the surface of a
cylinder of length~$L$ and circumference~$C$. By mapping it to a
ferromagnetic superspin chain, and using a scaling analysis, we show its
equivalence to a one-dimensional supersymmetric nonlinear sigma model in
the scaling limit, for any value of the ratio $L/C$, except for short
systems where $L$ is less than of order $C^{1/2}$. For the sigma model, the
universal crossover functions for the conductance and its variance have
been determined previously.  We also show that the DN model can be mapped
directly onto the random matrix (Fokker-Planck)  approach to disordered
quasi-one-dimensional wires, which implies that the entire distribution of
the conductance is the same as in the latter system, for any value of $L/C$
in the same scaling limit. The results of Chalker and Dohmen are explained
quantitatively.

\end{abstract}


\vskip -5mm

\section{Introduction}
\label{1}

Disordered conductors have been at the focus of experimental and
theoretical research for quite some time. Even properties of a single
electron in random potential are quite nontrivial. One of the challenging
problems still open in this area is the description of the transition
between the plateaus in the integer quantum Hall (QH) effect. Chalker and
Coddington~\cite{cc} introduced a network model to deal with this problem,
and studied it numerically. Later several authors mapped this model to an
antiferromagnetic spin chain, using replicas or supersymmetry to average
over the disorder~\cite{r,l,z,k}. The analysis of this spin chain is still
complicated, and this model is not solved at present.

Recently Chalker and Dohmen~\cite{cd} introduced a three-dimensional (3D) 
version of the network model to study the possibility of QH effect in 3D
conductors.  This 3D network models a conductor composed of stacked
coupled layers perpendicular to a strong magnetic field. Each layer may
separately exhibit the QH transition and is represented by the usual 2D
network. In this system there exists a phase in which each individual
layer is in the middle of a QH plateau, and all the electronic states in
the bulk are localized. The only current-carrying states are the edge
states on the surface of the conductor. These chiral surface states were
studied in~\cite{k,cd,bafi,bfz,m,y}. In the framework of the network
model, these surface states form a 2D {\it directed\/} network (DN),
equivalent to one studied before~\cite{skr}. A distinguishing feature of
the model studied by Chalker and Dohmen is the periodic boundary condition
on the edge of each layer, making the system the surface of a cylinder.

In this paper, we study (following Chalker and Dohmen) the conductance
properties along the axis of the cylinder in the DN model. The paper is
organized as follows. We set up a general formalism for the description of
the DN model using supersymmetry and the approach of Ref.~\cite{r} in
section~\ref{2}. Then we go to a continuum limit, and map the DN model to
a spin chain in section~\ref{3} (this was also done in
Refs.~\cite{k,bafi,bfz}). Unlike the case of the QH transition, this spin
chain is {\it ferromagnetic}. This allows us to analyze its properties
using ideas of a recent paper on continuum quantum ferromagnets~\cite{rs}.
Scaling arguments, similar to those of Ref.~\cite{rs}, show that all the
moments of the conductance and other observables behave universally in the
scaling limit (i.e. are given by universal scaling functions of
dimensionless combinations of the {\it bare\/} couplings of the continuum
model).

In particular, we show that when the ferromagnetic chain under
consideration is in the classical regime (called ``renormalized''
classical in Ref.~\cite{rs}, but we drop the modifier because, due to the
no-scale-factor universality~\cite{rs}, no coupling constants are
renormalized), it can be further reduced to a 1D classical non-linear
sigma model, studied before by Mirlin {\it et al}.\ in the context of
localization in quasi-1D wires~\cite{mmz}. Based on this reduction we show
that in this classical scaling regime all the moments of the
conductance~$g$ are the same for both models.  Thus, borrowing results of
Ref.~\cite{mmz}, we can fully determine the crossover functions for the
mean and variance of the conductance of the DN model for {\it any\/} ratio
of the length and circumference of the cylinder in the scaling regime.
This classical scaling regime does {\it not\/} include the regime of very
small $L < {\cal O} (C^{1/2})$ that was termed 0D in Ref.~\cite{bfz}. 

A recent paper~\cite{bf} shows how one can in principle obtain the full
probability distribution of the transmission eigenvalues from the
non-linear sigma model, and proves that this distribution is identical to
the one obtained from the Dorokhov-Mello-Pereyra-Kumar (DMPK) equation
\cite{d,mpk} of the Fokker-Planck (FP) approach which describes the
universal behavior of localization in quasi-1D wires. In view of this
equivalence, the probability distribution of the conductance of the DN
model in the classical scaling regime is the same as that of the quasi-1D
model.  For the quasi-1D model, a lot is known exactly about this
distribution~\cite{br}, and thus a nearly complete description of the
conductance properties of the DN model is available. In section~\ref{4} we
give a direct argument that shows how the DN model is related to a
quasi-1D model, and thus to the DMPK equation. We compare our results with
those of Chalker and Dohmen~\cite{cd}. We conclude in section~\ref{5}. 

\section{The Directed Network Model: General Setup and the Symmetry}
\label{2}

\begin{figure}
\epsfxsize=3.4in
\centerline{\epsffile{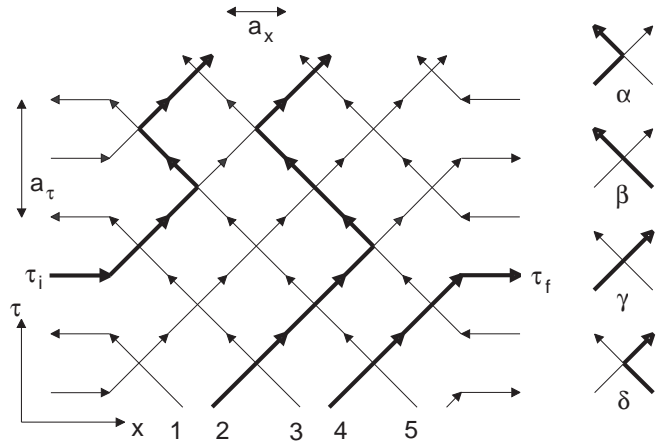}}
\vspace{0.25in}
\caption{The directed network (DN) model. The particles move on the links
in the directions shown by the arrows. The bold line represents a typical
path. The scattering amplitudes on such a path are as shown on the right.}
\label{fig1}
\end{figure}

The DN is shown in Fig.~\ref{fig1} and consists of links and nodes. The
links carry complex fluxes, and the nodes represent (unitary) scattering
matrices ${\cal S}$ connecting incoming $(i, i')$ and outgoing $(o, o')$
fluxes:
\be
\label{smat}
\left( \begin{array}{c} o \\ o' \end{array} \right) =
{\cal S} \left( \begin{array}{c} i \\ i' \end{array} \right) =
\left( \begin{array}{cc} \alpha & \beta \\
\gamma & \delta \end{array} \right)
\left( \begin{array}{c} i \\ i' \end{array} \right).
\ee
The scattering amplitudes $\alpha, \ldots, \delta$ correspond to
elementary scattering events shown on the right in Fig.~\ref{fig1}. For
the time being they are assumed to be arbitrary complex numbers different
for different nodes, which allows us to formulate our model for disordered
samples with any realization of disorder. A particular distribution for
the scattering amplitudes will be specified later.  The vertical direction
in Fig.~\ref{fig1} is along the circumference of the cylinder. Later it
will play the role of imaginary time for the spin chain, so we call the
vertical coordinate~$\tau$. In the $\tau$-direction the network has the
size $C=N_{\tau} a_{\tau}$ where $N_{\tau}$ is the number of ``channels''
through the system and $a_{\tau}$ is a microscopic scale of the order of
the mean free path of electrons. We impose periodic boundary conditions in
this direction. In the $x$-direction the network has finite length $L=N
a_x$, where~$N$ is the number of layers (or ``sites''), and $a_x$ is the
distance between them.  In this direction the system is connected to ideal
leads. In Fig.~\ref{fig1} the edge states are numbered from~1 to $N = 5$,
and $N_{\tau} = 3$.

Different correlation functions may be defined for this model and each of
them may be represented in the first or second quantized way. As an
illustrative and important example, we derive expressions for the
conductance. The dimensionless conductance is given by the Landauer
formula $g = \mbox{tr \boldmath $t^{\dagger}t$} = \sum_{i,f}
|t_{if}|^{2}$, where~{\boldmath $t$} is the total transmission matrix
(with matrix elements $t_{if}$)  between left and right boundaries of the
system. In the first quantized language, $t_{if}$ is given by the sum over
``retarded'' paths connecting an incoming link at $\tau_i$ on the left
boundary and an outgoing link at $\tau_f$ on the right boundary.  Each
path follows links only in the direction of the arrows and its
contribution is the product of the scattering amplitudes along the path.
One such path is shown on Fig.~\ref{fig1} with bold lines.  Similarly,
$t_{if}^{*}$ is given by the sum over ``advanced'' paths where each node
contributes the complex conjugate scattering amplitude.

In the second quantized language, the sum over paths may be written as a
trace of an ``evolution'' operator in a Fock space of bosons and
fermions. To represent $t_{if}$ ($t_{if}^*$) we introduce a retarded
(advanced) boson $a_i$ ($b_i$) and fermion $f_i$ ($g_i$) on each link~$i$.
The numbers of bits of paths on each link play the role of occupation
numbers of these bosons and fermions, and the collection of these numbers
on horizontal section at a given $\tau$-coordinate specifies a state in
the Fock space. Then the evolution operator $V_{12}$ for a single node
between the sites~1 and~2, which evolves quantum states on these sites by
one step in the $\tau$-direction, is given by the sum of the contributions
of all possible scattering events, described by ${\cal S}$.  In a typical
event involving only retarded bosons,~$k$ bosons are transferred from
site~1 to site~2 with amplitude~$\gamma$,~$l$ bosons are transferred from
site~2 to site~1 with amplitude~$\beta$, and the remaining bosons stay on
their respective sites, each contributing factors~$\alpha$ for site~1
and~$\delta$ for site~2. This event gives the term
$ (\gamma^k \beta^l \! /{k! \, l!}) (\ad_2)^k (\ad_1)^l
\alpha^{n_{a1}} \delta^{n_{a2}} a_2^l a_1^k$
(where $n_{a1}=\ad_1 a_1$ etc.) in the evolution operator $V_{12}$.  After
some rearrangement, the sum of all such terms for all the bosons and
fermions may be written as
\bea
\label{V}
V_{12} & = & : \exp \biggl(
{\gamma \over \alpha} (\ad_2 a_1 + \fd_2 f_1) +
{\beta \over \delta} (\ad_1 a_2 + \fd_1 f_2) \nonumber \\
& & + {\gamma^* \over \alpha^*} (\bd_2 b_1 + \gd_2 g_1) +
{\beta^* \over \delta^*} (\bd_1 b_2 + \gd_1 g_2) \biggr): \nonumber \\
& & \times \alpha^{n_{a1}+n_{f1}}
(\alpha^*)^{n_{b1}+n_{g1}} \delta^{n_{a2}+n_{f2}}
(\delta^*)^{n_{b2}+n_{g2}},
\eea
where colons stand for normal ordering.

\begin{sloppypar}
The contribution of the boundary nodes connected to the leads is
different.  When representing $|t_{if}|^2$, every leftmost node at $\tau
\neq \tau_i$ has only paths reflected off the left boundary (with the
corresponding amplitude~$\delta$ or~$\delta^*$), because only such paths
contribute to $|t_{if}|^{2}$.  Then the only possible event at such a node
is that all the particles stay on the site~1, each contributing
factors~$\delta$ for retarded and~$\delta^*$ for advanced ones. The
corresponding evolution operator is simply
\be
V_{01} = \delta^{n_{a1} + n_{f1}} (\delta^*)^{n_{b1} + n_{g1}}.
\ee
For the leftmost node at $\tau_i$ we also need to inject one retarded and
one advanced path into the system.  This is represented by the event where
we create additional retarded and advanced particles on the site~1. For
definiteness we choose them to be bosons. The corresponding evolution
operator is $|\gamma|^2 \ad_1 \bd_1 \delta^{n_{a1} + n_{f1}}
(\delta^*)^{n_{b1} + n_{g1}} = |\gamma|^2 \ad_1 \bd_1 V_{01}$. Similarly,
for the rightmost nodes at $\tau \neq \tau_f$ we have
\be
V_{N,N+1} = \alpha^{n_{aN} + n_{fN}} (\alpha^*)^{n_{bN} + n_{gN}},
\ee
and for the boundary node at $\tau_f$ the evolution operator is
$|\gamma|^2 V_{N,N+1} a_N b_N$.
\end{sloppypar}

\begin{sloppypar}
The total evolution operator~$U$ is composed of all the $V_{i,i+1}$'s in
the following manner. Assume for definiteness that the number of
sites~$N$ is odd. Then in our system we have even rows of links at
integer times~$0, a_{\tau}, 2 a_{\tau}$, etc., where fluxes enter the
system from the left and exit it to the right, and odd rows of links at
half-integer times $a_{\tau}/2, 3 a_{\tau}/2$, etc., where fluxes
enter from the right and exit to the left. For even rows we form
the product
\be
V_e(\tau) = V_{01}(\tau) V_{23}(\tau)  \ldots V_{N-1,N}(\tau),
\ee
and for odd ones
\be
V_o(\tau) = V_{12}(\tau) V_{34}(\tau)  \ldots V_{N,N+1}(\tau).
\ee
The operator~$U$ is then given by the product
\be
U = \prod_{n=0}^{N_{\tau}-1} V_o\Bigl( (n+1/2)a_{\tau}\Bigr)
V_e\Bigl( n a_{\tau}\Bigr),
\ee
which is ordered with the earliest times at the right. Note that the
only $\tau$-dependence in the operators $V_e$, $V_o$ is through the ${\cal
S}$-matrix elements, which implicitly depend on $i$ and $\tau$.

\end{sloppypar}

With the help of the operator~$U$, the conductance is given by
\be
\label{cond}
g = {\rm STr} \biggl( {\cal T} \! \sum_{\tau_i,\tau_f} |\gamma_i|^2
\ad_1(\tau_i) \bd_1(\tau_i) |\gamma_f|^2 a_N(\tau_f) b_N(\tau_f) \, U
\biggr) \!.
\ee
Here $\tau_i$ and $\tau_f$ label the times at which the creation and
annihilation operators act on the states, $\gamma_i$ and $\gamma_f$ are
the ${\cal S}$ matrix elements at the corresponding nodes, and~${\cal T}$
orders times, placing the earliest at the right. ``STr'' stands for the
supertrace in the Fock space, which weights all the states with the factor
of $(-1)^{N_F}$, where $N_F = \sum_{i=1}^N (n_{fi} + n_{gi})$ is the total
number of fermions in a state, ${\rm STr} (\ldots) = {\rm Tr} ((-1)^{N_F}
\ldots)$. That is, states with an odd number of fermions contribute to
the sum with a negative sign.  This, together with the periodicity in the
$\tau$-direction, ensures the cancellation of bosonic and fermionic
contributions from closed paths not connected to the leads.  These closed
paths were, as usual, absent from the original first quantized formulation
of the problem.

In Eq.~(\ref{cond}), $\tau_i$ and $\tau_f$ take only half-integer
values $a_{\tau}/2$, $3a_{\tau}/2$, so on, because we create and destroy
bosons only on half of the links belonging to the sites~1 and~$N$.
Using the commutation relations between the bosonic operators and~$V$'s,
we can rewrite the expression for~$g$ in the form
\be
\label{cond1}
g = {\rm STr} \biggl( {\cal T}
\sum_{\tau} \varepsilon_{\tau} \ad_1(\tau) \bd_1(\tau)
\sum_{\tau'} \varepsilon_{\tau'} a_N(\tau') b_N(\tau') \, U \biggr) \!,
\ee
where now $\tau$ and $\tau'$ take all possible values, but
$\varepsilon_{\tau} = 1$ ($-1$) for integer (half-integer) values of
$\tau/a_{\tau}$.

This is a good place in which to discuss in more detail the choice of
boundary conditions for our model. In the $x$-direction the presence of
the ideal absorbing leads at the boundaries of the system means in the
first quantized language that we do not include contributions from the
paths leaving or entering the system unless we specifically calculate some
correlators between the boundaries (like conductance~$g$). In the second
quantized language this translates to the following. We can imagine having
two additional vertical sets of links in the leads~(0th and $N+1$st sites)
on which we have no bosons or fermions, so these sites always carry the
vacuum state $| 0 \rangle$. With this constraint the boundary operators
$V_{01}$ and $V_{N,N+1}$ are seen to be special forms of general
$V_{i,i+1}$'s, Eq.~(\ref{V}), acting on the vacuum at the left or right.
Later this constraint on the states at the boundaries will give, in the
$\tau$-continuum formulation, the symmetry-breaking term in the
Hamiltonian and in the action, and will fix the boundary conditions in the
continuum field theory, see Eq.~(\ref{bc}) below.

We now discuss the supersymmetry properties of our formulation.
For each site~$i$ we can form~16 bilinears in our bosonic and fermionic
operators. They represent the~16 generators of the Lie superalgebra
$u(1,1\,|\,2)$.  We arrange these generators in a~$4 \times 4$ matrix, or
``superspin''~$J_i$:
\be
J_i= \!\! \left( \!\! \begin{array}{cccc}
a_i \ad_i - 1/2 & a_i \fd_i & a_i b_i & a_i g_i \\
f_i \ad_i & f_i \fd_i - 1/2 & f_i b_i & f_i g_i \\
-\bd_i \ad_i & -\bd_i \fd_i & -\bd_i b_i - 1/2 & -\bd_i g_i \\
\gd_i \ad_i & \gd_i \fd_i & \gd_i b_i & \gd_i g_i-1/2
\end{array} \!\! \right) \! .
\ee
We adopted ``advanced-retarded'' arrangement of the generators rather than
the ``boson-fermion'' one (see~\cite{z} for details). In this scheme the
diagonal~$2 \times 2$ blocks of $J_i$ contain the generators of the
subalgebra $u(1\,|\,1) \oplus u(1\,|\,1)$. Then one can show that {\it the
operator\/} $V_{12}$ {\it commutes with all~16 components of\/} $J_1+J_2$
({\it and thus has\/} $u(1,1\,|\,2)$ {\it as the symmetry algebra{\rm)}},
\be
\label{susy}
(J_1 + J_2) \, V_{12} = V_{12} \, (J_1 + J_2),
\ee
{\it if and only if the scattering matrix\/}~${\cal S}$ {\it is
unitary\/}: ${\cal S}^{-1} = {\cal S}^{\dagger}$ (the analogous result for
the Chalker-Coddington model was found in Ref.~\cite{r}). Thus, the
products of the $V_{i,i+1}$'s in $V_e$ and $V_o$ commute with the total
superspin $\sum_{i=1}^N J_i$ for any realization of disorder, provided
that all the scattering matrices are unitary, except for the boundary
operators $V_{01}$ and $V_{N,N+1}$. These boundary operators commute only
with generators from diagonal blocks of $J_1$ and $J_N$, correspondingly,
and break the symmetry of the total evolution operator~$U$ down to
$u(1\,|\,1)  \oplus u(1\,|\,1)$. Therefore, for any realization of the
disorder, unitarity of the scattering matrices ensures the global
$u(1,1\,|\,2)$ symmetry of our problem, which is broken only by the
boundary constraint.  The use of the supertrace, not the ordinary trace,
is essential in maintaining the supersymmetry in the presence of the
periodicity in the $\tau$-direction. 

In terms of the superspins, Eq.~(\ref{cond1}) for the conductance may be
rewritten as
\be
\label{cond2}
g = {\rm STr} \biggl( {\cal T} \sum_{\tau}
\varepsilon_{\tau} J_1^{31}(\tau)
\sum_{\tau'} \varepsilon_{\tau'} (-1)^N J_N^{13}(\tau')
\, U \biggr) \!.
\ee
Here $J_1^{31} = - \bd_1 \ad_1, J_N^{13} = a_N b_N$ are particular
components of superspins $J_1$ and $J_N$ at the boundaries of the system.
The alternating sums of the type
\be
\label{curr1}
{\cal I}_i = (-1)^i \sum_{\tau} \varepsilon_{\tau} J_i^{ab}(\tau)
\ee
represent the total current through the system in the $x$-direction.
Using the supersymmetry of the operators~$V$, Eq.~(\ref{susy}), we can
show that this current is conserved, i.e., correlators of ${\cal I}_i$'s
do not depend on the position labels $i$, except at the boundaries. We see
then that Eq.~(\ref{cond2}) is the usual Kubo-type formula, relating the
conductance to a current-current correlator, $g = - {\rm STr} ({\cal T}
\,{\cal I}_i {\cal I}_j U)$, for $i \neq j$.

Similarly we can write down expressions for the moments of the conductance
or other correlators. For moments higher than the second, we need to
introduce some additional structure. Namely, we introduce $n$ replicas of
our bosons and fermions and sum the bilinear operators in the exponentials
in the expressions for $V_{i,i+1}$'s over replica indices. This is
necessary, because to represent the $m$-th power of the conductance we
need to create at the boundary~$m$ {\it different\/} retarded and~$m$ {\it
different\/} advanced particles, and we need $n \geq m/2$. The
corresponding model has global $u(n,n\,|\,2n)$ symmetry broken down to
$u(n\,|\,n) \oplus u(n\,|\,n)$ by the boundary constraint. We should point
out that in this case replicas are {\it not\/} introduced to average over
the disorder, and their number~$n$ is {\it not\/} taken to zero in the
end.  In this model, results for the $m$-th moment are independent of $n$,
provided $m \leq 2n$, because any ``excess'' replicas cancel by
supersymmetry. For simplicity we continue to assume $n=1$.

So far we did not have to specify the nature of the disorder. Now we
assume a particular distribution of the scattering matrices, used before
in~\cite{cc,cd}. Namely, we take every scattering matrix to consist of a
product of two diagonal unitary matrices with a real orthogonal matrix,
with entries
\be
\label{scampl}
-\beta = \gamma = t > 0, \quad \alpha = \delta = (1-t^2)^{1/2},
\ee
in between them, the latter matrix being the same for all nodes.  The
phases from the diagonal unitary matrices are associated with links rather
than with nodes and are assumed to be uniformly distributed between~0
and~$2\pi$, independently for each link~\cite{cc,cd}. Thus averaging over
disorder corresponds to integration over all the link phases.

It is easy to see that this averaging produces a local constraint on the
states in the Fock space. Let each retarded bit on a given link at the
$i$-th site at time~$\tau$ contribute the factor $e^{i\varphi}$, and each
advanced one the factor $e^{-i\varphi}$, to the amplitude associated with
a given path. Then if we have~$m$ retarded and~$n$ advanced bits on the
link, averaging over random phase $\varphi$ gives $\int_0^{2\pi} \exp
\bigl( i(m-n) \varphi \bigr)  d\varphi = \delta_{m,n}$, that is, we get
zero unless the number of retarded and advanced bits is the same. Thus, we
have a local constraint on the number of bits of paths on each link. In
the second-quantized language this means that the averaging projects our
evolution operator to the subspace specified by $n_{ai} + n_{fi} = n_{bi}
+ n_{gi}$ for each~$i$ and each time~$\tau$. This subspace forms a highest
weight irreducible representation of the algebra $u(1,1\,|\,2)$, with the
vacuum $| 0 \rangle$ for the site $i$ being the ``highest weight'' vector.
This representation was first obtained for the Chalker-Coddington
model~\cite{cc} in~\cite{r} and was also discussed in~\cite{z}.

\section{Mapping to a Spin Chain and Scaling Analysis}
\label{3}

So far in our derivation we retained discreteness of the DN. Unlike in the
other published derivations~\cite{k,bafi,bfz}, we did not have to take the
time-continuum limit from the start. However, we do so now in order to
obtain a continuum field theory for our system and to study its universal
scaling properties. In the $\tau$-continuum limit we assume $t \ll 1$ and
expand the $V_{i,i+1}$'s, projected to the constrained subspace, to second
order in~$t$. The result of the expansion is $V_{i,i+1} = 1 - t^2 {\rm
str} J_i J_{i+1} + {\cal O}(t^4)$, where $J_i$ are superspins in the
highest weight representation mentioned above. Here ``str'' stands for the
matrix supertrace in the space labeled by the upper indices of $J_i^{ab}$.
In other words, for any $4\times 4$ supermatrix~$A$, str$A \equiv {\rm tr}
\, \eta A$ where ``tr'' is the usual matrix trace and~$\eta$ is a diagonal
matrix with entries $(1,-1,1,-1)$. Now we combine all the~$V$'s and
reexponentiate. We also replace sums over~$\tau$ by integrals. The result
is the evolution operator in imaginary time~$\tau$ of a 1D quantum
ferromagnetic spin chain with Hamiltonian
\be
H[J] = {t^2 \over a_{\tau}} \biggl( \sum_{i=1}^{N-1} {\rm str} J_i
J_{i+1} + {1 \over 2} {\rm str} \Lambda (J_1 + J_N) \biggr).
\ee
Here~$\Lambda$ is the diagonal matrix ${\rm diag}(1,1,-1,-1)$. The last
term in~$H$ comes from the boundary nodes.  It may be interpreted as saying
that there are, at the boundaries, two additional spins $J_0$ and
$J_{N+1}$ fixed to a particular ``direction'' $J_0 = J_{N+1} = \Lambda/2$
by an infinitely strong magnetic field coupling via a Zeeman term in the
superspin space.  This is another manifestation of the boundary constraint
mentioned above.

Using the algebra of the generators $J^{ab}$ in the same
$\tau$-continuum limit, the expression~(\ref{curr1}) for the
total current becomes
\be
\label{curr2}
{\cal I}_i = {t^2 \over a_{\tau}} \int_0^C \!\! d\tau [J_i, J_{i-1}] =
- {t^2 \over a_{\tau}} \int_0^C \!\! d\tau [J_i, J_{i+1}].
\ee
For the particular components of the current at the boundaries these
expressions reduce to ${\cal I}_1^{31} = (t^2/2a_{\tau}) \int \! d\tau
[J_1, \Lambda]^{31} = (t^2/a_{\tau}) \int \! d\tau J_1^{31}$ and,
similarly, ${\cal I}_N^{13} = (t^2/a_{\tau}) \int \! d\tau J_N^{13}$.

Next, we introduce supercoherent states and represent quantities of
interest as path integrals over some supermanifold, see~\cite{bfz,rs1} for
details. The resulting theory has the action 
\be 
S = S_B + \int_0^C \!\! d\tau H[Q(\tau)], 
\ee 
where $S_B$ is the Berry phase term (specified below), $Q_i(\tau)$ is a
supermatrix taking values in the coset space $U(1,1\,|\,2)/U(1\,|\,1) 
\times U(1\,|\,1)$ (where $U(1,1\,|\,2)$ is the supergroup of which the
Lie superalgebra is $u(1,1\,|\,2)$, etc.), and $H[Q(\tau)]$ is obtained
from $H[J]$ by replacing every $J_i$ with $Q_i(\tau)/2$ (we make the same
replacement in the expressions for the total current). In the path
integral, all the components of $Q$ obey periodic boundary conditions. The
difference, in the case of the fermionic components, from the usual path
integrals for fermions, which obey antiperiodic boundary
conditions~\cite{no}, is a direct consequence of the factor $(-1)^{N_F}$
in the definition of STr. 

The scaling properties of quantum ferromagnets were discussed
in~\cite{rs}.  Following that paper, we take the spatial continuum limit
of the action~$S$ and the current ${\cal I}_i$. The resulting continuum
ferromagnet has the action
\bea
\label{scont}
S_{\rm cont} & = & - \int_0^L \!\! dx \!\! \int_0^C \!\! d\tau \biggl(
\! {M_0 \over 2} \!\! \int_0^1 \! du \, {\rm str} \, Q(u) {\partial Q(u)
\over \partial u} {\partial Q(u) \over \partial \tau} \nonumber \\ & & +
{\rho_s \over 2} {\rm str} (\nabla Q)^2 \biggr) \!,
\eea
and the current becomes
\be
{\cal I}(x) = -2 \rho_s \int_0^C \!\! d\tau Q \nabla Q.
\ee
Here $M_0 = 1/2 a_x$ and $\rho_s = t^2 a_x/4 a_{\tau}$ are the
magnetization per unit length, and the spin stiffness in the ferromagnetic
ground state respectively. Also, in the Berry phase term, the first term in
$S_{\rm cont}$, $Q(u) \equiv Q(x,\tau,u)$ is some smooth homotopy between
$Q(x,\tau,0) = \Lambda$ and $Q(x,\tau,1) = Q(x,\tau)$ (for details
see~\cite{rs1}). The circumference of the cylinder~$C$ plays the role of
the inverse temperature. The last term in $H[Q(\tau)]$, or the
interpretation $J_0 = J_{N+1} = \Lambda/2$, forces the field $Q(x,\tau)$
to take the boundary values
\be
\label{bc}
Q(0,\tau) = Q(L,\tau) = \Lambda.
\ee
A boundary condition of this form to represent ideal absorbing leads is
usual in the non-linear sigma model formulation of the theory of
localization (to which, we note, the model~(\ref{scont}) is {\em not}
equivalent). Given that the effective supersymmetric spin system is
ferromagnetic, and that fluctuations in the length of the spins can be
neglected, the form of the action $S_{\rm cont}$ is dictated by the
supersymmetry, which holds universally for any short range form of
disorder.

The action~(\ref{scont}) clearly shows an anisotropic scaling which
reflects the difference between the $x$- and $\tau$-directions in the DN.
In the original discrete version of the DN model, if we consider the local
behavior without the periodic boundary condition, the mean conductivity
behaves diffusively in the $x$-direction, with $\tau$ playing the role
of time, with the diffusion constant
\be
\label{diff}
D = {t^2 \over 1-t^2} {a_x^2 \over a_{\tau}},
\ee
as shown in Refs.~\cite{cd,skr}. (However, even without the periodic boundary
condition, less trivial behavior is found for higher moments~\cite{skr}.)
In the spin chain language this anisotropy is manifest in the usual
quadratic dispersion relation of the spin waves. The action~(\ref{scont}),
linearized near the ferromagnetic ground state, describes (on analytic
continuation to real time) spin waves with the dispersion $\omega =
2\rho_s k^2/M_0$. The difference between the ratio $2\rho_s/M_0 = t^2
a_x^2/a_{\tau}$ and the diffusion constant $D$ is due to the particular
order of time- and space-continuum limits taken to obtain the
action~(\ref{scont}). The more general long wavelength behavior ($k_x a_x
\ll1$, $k_{\tau}a_{\tau} \ll 1$), consistent with Eq.~(\ref{diff}), should
be described by the same action $S_{\rm cont}$ but with the more accurate
expression for the spin stiffness,
\be
\rho_s = {t^2 \over 1-t^2} {a_x \over 4a_{\tau}}.
\ee
In general, the continuum description requires that $C \gg a_{\tau}$, $L
\gg a_x$. 

Although $S_{\rm cont}$ is a highly non-trivial interacting quantum field
theory, it is nevertheless possible to make some simple exact statements
about it. First, its ``ground state'' (which dominates the functional
integral in the limit of zero ``temperature'', $C=\infty$) is fully
polarized and fluctuationless, so $\langle Q \rangle_{C=\infty} =
\Lambda$. As a result, the parameter $M_0$, which measures the Berry phase
due to adiabatic changes in the ground state polarization, cannot have any
non-trivial zero temperature renormalization~\cite{rs}. Second, not only
the ground state, but some low-lying excited states are also known
exactly: the linearized {\it single\/} spin-wave states are in fact exact
eigenstates of the full Hamiltonian, and their dispersion $\omega = 2
\rho_s k^2 / M_0$ has no corrections from the non-linearities. This
implies that $\rho_s$ also has no non-trivial zero-temperature
renormalization~\cite{rs}. We also mention, as an aside, that the mapping
from the DN model to $S_{\rm cont}$ can be generalized to higher spatial
dimensions $d$. As a result, the critical dimension $d=2$ (for the
properties at large $C$) pointed out in Ref.~\cite{rs} is related to the
same fact obtained in Ref.~\cite{skr}.

The absence of any non-trivial renormalizations of $M_0$ and $\rho_s$, the
absence of any additional relevant operators in $1+1$ dimensions, and the
related absence of ultraviolet divergences in physical quantities in the
$1+1$-dimensional quantum field theory, are responsible for the phenomenon
of ``no-scale-factor universality''~\cite{rs,sss}.  Among its implications
is that scaling analysis of $S_{\rm cont}$ reduces to the naive analysis
of engineering dimensions.  The field $Q(x,\tau)$, being subject to the
constraint $Q^2(x,\tau)=1$, is dimensionless. Then simple power counting
gives the engineering dimensions of the couplings: ${\rm dim}\,M_0 = ({\rm
length})^{-1}$, ${\rm dim}\,\rho_s = {\rm length/time}$.  The
no-scale-factor universality also implies that all the observables are
functions of dimensionless combinations of the bare couplings $M_0$ and
$\rho_s$ and the large scales, which, for the conductance, are just $L$
and $C$. For example, the mean conductance is given by a function
\be
\mean{g} = \Phi_g \left( \frac{L}{(C \rho_s / M_0)^{1/2}}, 
\frac{L}{C \rho_s} \right).
\label{scfun}
\ee
Clearly, we could have chosen other combinations of the two arguments of
the $\Phi_g$, and the physics behind our particular choices will become
clear as we proceed. The continuum theory $S_{cont}$ applies to any
particular lattice quantum ferromagnet provided the lengths $L,C$ are
sufficiently large. Specifically~\cite{rs}, we require $M_0 L \gg 1$ and
$\rho_s M_0 C \gg 1$ (or $L \gg a_x$ and $C \gg a_{\tau} (1-t^2)/t^2$, as
well as $C \gg a_{\tau}$).  Comparing with the function $\Phi_g ( \ell_1 ,
\ell_2 )$, we see that the condition on $C$ implies that $\ell_1 \gg
\ell_2$, and $\Phi_g$ becomes universal in this regime, with no
non-universal factors in either the scale of $\Phi_g$ or its arguments.
Notice that the arguments $\ell_1$, $\ell_2$ can take all positive value
$0 < \ell_1, \ell_2 \leq \infty$ while satisfying $\ell_1 \gg \ell_2$. We
also emphasize that the condition for universality may be satisfied for
any ratio $L/C$, {\em i.e.\/} for short as well as long cylinders.

\begin{figure}
\epsfxsize=3.4in
\centerline{\epsffile{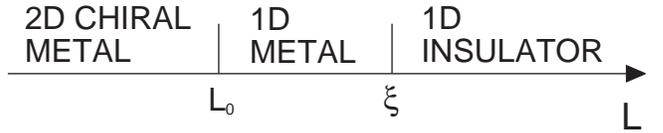}}
\vspace{0.2in}
\caption{Sketch of the crossovers as a function of the length $L$. The
first crossover~\protect\cite{bfz} is at the scale $L \sim L_0 = (C \rho_s
/ M_0 )^{1/2}$ where the low-lying energy level splittings of the quantum
ferromagnet $\sim \rho_s / M_0 L^2$ are of order the ``temperature''
$1/C$.  The second crossover is near the localization length $\xi = 8 C
\rho_s$.  The conditions for the validity of the continuum theory require
$\xi \gg L_0$, and so the two crossover scales are well separated. The
quantum ferromagnet behaves classically for all $L \gg L_0$ (but with no
restriction on $L/C \rho_s$) and is then described by the one-dimensional
supersymmetric non-linear sigma model $S_1$ (Eq.~(\protect\ref{ss2})),
which is therefore valid across the crossover at $L \sim C \rho_s$. The
Fokker-Planck approach of Section~\protect\ref{4} has the same regime of
validity ($L \gg L_0$, $L/C \rho_s$ arbitrary). The 2D chiral metal regime
was labelled 0D in Ref.~\protect\cite{bfz}, but we prefer the present
terminology for reasons discussed below Eq.~(\protect\ref{ss3}). }
\label{fig2}
\end{figure}

We first present our results for the case $L \gg L_0 \equiv (C \rho_s /
M_0 )^{1/2}$. The length $L_0$ was identified in Ref.~\cite{bfz}, and we
will discuss the regime $L \leq L_0$ later (see Fig~\ref{fig2}). For $L
\gg L_0$, we have from (\ref{scfun}) 
\be
\label{scfun1}
\mean{g} = \Phi_{g1} \left({L \over C \rho_s} \right),
\ee
where $\Phi_{g1}( \ell )  \equiv \Phi_g (\infty, \ell)$. Notice that the
ratio $L/C$ is still allowed to be arbitrary.  The form~(\ref{scfun1})
implies that any localization length $\xi$ must obey $\xi \propto
C\rho_s$, which was the basic result of Chalker and Dohmen~\cite{cd},
which we have now explained. One might have been tempted to use the
incorrect naive argument, that distances in the $x$-direction (such
as~$\xi$) should scale as the square root of distances in the
$\tau$-direction (such as $C$).  However, under such scaling, $\rho_s$ is
not scale invariant, while $C\rho_s$ {\em does} scale as a length. In the
underlying continuum quantum ferromagnet, there are two dimensionful
parameters $\rho_s$ and $M_0$, so this is not a scale-invariant system
either, unlike, for example, the usual 2D non-linear sigma model at the QH
critical point. 

It is possible to obtain scaling functions like $\Phi_{g1}$ directly
from a certain field theory. They are current-current correlators in the
path integral with the 1D action
\be
S_1 = - {C \rho_s \over 2} \int_0^L \! dx \, {\rm str} (\nabla Q)^2,
\label{ss2}
\ee
obtained from $S_{\rm cont}$ by neglecting the $\tau$-depen\-dence
of~$Q$~\cite{tns}. The contributions of all modes with a non-zero 
``frequency'' along the
$\tau$ direction to the coefficient of $(\nabla Q)^2$ in $S_1$ are suppressed by
powers of $L_0 / \xi$ and $L_0/L$.
Eqn (\ref{ss2}) is exactly the action of the 1D non-linear sigma
model studied by Mirlin {\it et al}. in Ref.~\cite{mmz}. In that paper, the
authors used harmonic analysis on superspaces to diagonalize the transfer
matrix of this 1D model. As a result they were able to express the mean
conductance $\mean{g}$ and its variance var\,$g$ as rather simple
integrals/sums over spectral parameters. For example,~$\mean{g}$ is given
by
\be
\label{cond4}
\mean{g} = 2 \!\! \sum_{\scriptstyle m > 0 \atop \scriptstyle {\rm odd}}
\! \int_0^{\infty} \!\!\!\!\! d\lambda \, \tanh{\pi \lambda \over 2} {m
\lambda \over m^2+\lambda^2} \exp \left( \! - {(m^2+\lambda^2) L \over 2
\xi} \right) \!,
\ee
where $\xi$ is the localization length,
\be
\label{loclength}
\xi = 8 C \rho_s = C {2 t^2 \over 1-t^2}{a_x \over a_{\tau}}.
\ee
(Note that we do not include any spin degeneracy, unlike Ref. \cite{mmz}.)

In the 1D metallic limit of short cylinders, $L_0 \ll L \ll \xi$,
Eq.~(\ref{cond4}) reduces to~\cite{cd}
\be
\label{metg}
\mean{g}_{L \ll \xi} = {\xi \over 2L} = {C \over L} \, {t^2 \over 1-t^2}
\, {a_x \over a_{\tau}},
\ee
while in the opposite 1D localized limit of long cylinders, $L \gg \xi$,
one
obtains
\be
\mean{g}_{L \gg \xi} = 2 \left(\pi \xi \over 2L\right)^{3/2} 
\exp\left(-{\displaystyle L \over \displaystyle 2\xi} \right).
\ee
An expression similar to Eq.~(\ref{cond4}) can also be written for the
variance ${\rm var} \, g$, see Ref. \cite{mmz}. In the metallic regime
$L_0 \ll L \ll \xi$, this yields the known result for universal
conductance fluctuations in 1D, ${\rm var} \, g = 1/15$.

We now turn to a discussion of the regime $L \leq L_0$ (Fig~\ref{fig2}). 
Different behavior arises here because~\cite{bfz}, in the ferromagnetic
language, the ``temperature'' $1/C$ becomes of order the low-lying level
splittings, which are of order $\rho_s / M_0 L^2$. In contrast, the
classical regime we have discussed above is where the temperature is much
greater than these quantum mechanical splittings. The crossover from the
very short regime, $L \ll L_0$ (termed 0D in Ref~\cite{bfz}), to the 1D
metallic regime, $L_0 \ll L \ll \xi$, will be described by the $L/C \rho_s
\rightarrow 0$ limit of scaling forms like (\ref{scfun}).  The scaling
forms can be expanded as a perturbation series in increasing powers of
$L/C \rho_s$, times a universal function of $L/L_0$ in each term. For
$\mean{g}$, we can show that this takes the form
\be
\mean{g} = {4 C \rho_s \over L} + {L \over C \rho_s} \Upsilon \biggl(
\frac{L}{L_0} \biggr) + {\cal O} \biggl(\biggl({L \over C \rho_s}\biggr)^2
\biggr).
\label{ss1}
\ee
To leading order in $L/C \rho_s$ there is no dependence on $L/L_0$, and
the result (\ref{metg}) is valid in both the 0D and the 1D metallic
regimes. The possible term of order $(L/C\rho_s)^0$ vanishes identically,
consistent with the known result that the leading ``weak localization''
correction to $\mean{g}$ vanishes in the quasi-1D metallic regime $L/L_0
\to \infty$ in the present (unitary) case.  The next term in the expansion
in $L/C\rho_s$ in (\ref{ss1}) does have a non-trivial crossover at the
scale $L_0$, described by the universal function $\Upsilon$, which should
approach the known quasi-1D metallic result $-1/180$ as $L/L_0 \to
\infty$~\cite{mmz}. In contrast, for the variance of $g$ we expect
\be
{\rm var} \, g = \widetilde{\Upsilon} \biggl(\frac{L}{L_0} \biggr) + 
{\cal O} \biggl(\frac{L}{C\rho_s}\biggr),
\label{ss3}
\ee
where the universal crossover from the 0D to the 1D metallic regime is now evident
in the leading term $\widetilde{\Upsilon}$, which must tend to 1/15 as
$L/L_0 \to \infty$. A result of this type, for ${\rm var} \, g$, has been
claimed by Mathur~\cite{m} and Yu~\cite{y}, who find ${\rm var} \, g
\propto C \rho_s/M_0 L^2$ in the 0D limit $L \ll L_0$.  This result
resembles that in isotropic systems, for example in 2D in the metallic
regime $\mean{g} \gg 1$, where ${\rm var} \, g \propto W/L$ (where $W$ is
the width), as $W/L \to \infty$ (see the results in Ref.~\cite{lsf}). Thus
a better name for this regime in the DN model would be 2D chiral metal,
and the crossover at $L_0 = (C \rho_s/ M_0)^{1/2}$ is from 2D to 1D
behaviour.

\section{Fokker-Planck Approach and Comparison to Earlier Work}
\label{4}

The 1D non-linear sigma model considered in the previous section is well
suited for obtaining moments of the conductance, but not for its
distribution. However, as we pointed out in the introduction and as we now
discuss, this 1D model is completely equivalent to the FP approach to the
conductance of quasi-1D wires in the limit of infinite number of channels
(``thick wire limit''), see Ref.~\cite{bf}. Then we can use the results of this
approach to obtain further properties of the DN model. In the FP approach one
concentrates on the eigenvalues $T_n = (\cosh x_n)^{-2}$ of the
transmission matrix {\boldmath $t^\dagger t$}.  The probability
distribution of the parameters $x_n$ satisfies the DMPK equation, which
was solved approximately in the localized and metallic regimes (see, for
example, Ref.~\cite{mc}) and exactly (for the unitary case) in
Ref.~\cite{br}. We summarize some of the results of this solution.

In the metallic limit, $L \ll \xi$, the $x_n$ have statistical
fluctuations, but the mean density of $x_n$'s is uniform, and the mean
conductance is equal to this density. This may be rephrased by saying that
the parameters $x_n$ are equally spaced in the average, $(x_n)_{L \ll \xi}
= n (x_1)_{L \ll \xi}$, and the mean conductance is equal to the inverse
of the first parameter $\mean{g}_{L \ll \xi} = (x_1)_{L \ll \xi}^{-1} =
\xi/2L$.

In the opposite localized limit, $L \gg \xi$, the parameters $x_n$ are
self-averaging (with normally distributed fluctuations), and their mean
positions are again equally spaced, but they are offset from the origin by
a half of the spacing between them:  $\mean{x_n}_{L \gg \xi} = (n-1/2)
(2L/\xi)$. In the localized limit, the conductance is dominated by the
smallest parameter, $g \approx 4 \exp(-2 x_1)$, and therefore its
logarithm is normally distributed with
\be
\label{logg}
\mean{\ln g}_{L \gg \xi} = - 2 \mean{x_1}_{L \gg \xi} = - {2L \over \xi}.
\ee
The product of $\mean{g}_{L \ll \xi}$ and $\mean{\ln g}_{L \gg \xi}$ is a
universal number:
\be
\label{univnum}
\mean{g}_{L \ll \xi} \, \mean{\ln g}_{L \gg \xi} = -1,
\ee
which is characteristic of the universal crossover from metallic to
localized behavior. If we now introduce rescaled parameters as
in~\cite{cd}, $\nu_n = (C/L) x_n$, then $\mean{\nu_1}_{L \gg \xi} =
C/\xi$, which for a given value of $C$ differs by a factor of $1/2$ from
$(\nu_1)_{L \ll \xi} = 2C/\xi$.

The equivalence, mentioned above, of the 1D sigma model and the FP
approach motivated us to look for a direct mapping from the DN model to
the FP equation, and thus for a better understanding of why the DN model
behaves as a quasi-1D conductor. For this purpose instead of considering
the evolution operators $V_e$ and $V_o$, which are the transfer matrices
in the $\tau$-direction (or ``row transfer matrices'') in the second
quantized language, we should concentrate on the transfer matrices in the
$x$-direction, or ``column transfer matrices''. For a single node such a
transfer matrix ${\cal M}$ connects fluxes on the left $(i, o)$ and on the
right $(i', o'$) of the node and is given by
\be
\label{tmat}
\left( \begin{array}{c} o' \\ i' \end{array} \right) =
{\cal M}  \left( \begin{array}{c} o \\ i \end{array} \right) =
\left( \begin{array}{cc} \delta/\beta & 1/\gamma^* \\
1/\beta & -\alpha/\beta \end{array} \right)
\left( \begin{array}{c} o \\ i \end{array} \right)
\ee
with $\alpha, \ldots, \delta$ being the same as in the Eq.~(\ref{smat}).
After going from one site to the next through one column of nodes the
directions of the fluxes are reversed at each $\tau$-coordinate. Then the
natural transfer matrices $M_{i-1,i+1}$ (which can be simply multiplied)
are composed of all the ${\cal M}$'s for two adjacent columns of the nodes
to the left and right of the $i$-th edge state. The total transfer matrix
for the DN with an odd number of edge states $N$ is then given by $M_N =
M_{N-1,N+1} \cdots M_{2,4}M_{0,2}$. This matrix connects
$N_{\tau}$-dimensional vectors of fluxes on the right of the system $(I',
O')$ with the ones on the left $(I, O)$:
\be
\label{Tmat}
\left( \begin{array}{c} O' \\ I' \end{array} \right) =
M_N  \left( \begin{array}{c} I \\ O \end{array} \right) =
\left( \begin{array}{cc} m_1 & m_2 \\
m_3 & m_4 \end{array} \right)
\left( \begin{array}{c} I \\ O \end{array} \right).
\ee

Now we neglect all the link phases (we will reinstate them later) and
consider the limiting case $\alpha = \delta = 0$ and $\gamma = - \beta
=1$, where back scattering from right- to left-moving flux, or vice versa,
is absent. In this limit, the column transfer matrix $M_{i-1,i+1}$ becomes
the ``shift'' matrix, which means that all the right- (left-) moving
fluxes are transferred without any change along the $x$-direction by
$2a_x$, and along the $\tau$-direction by $a_{\tau}$ ($-a_{\tau}$). In
other words, $M_{i-1,i+1}$ cyclically shifts all the right-moving fluxes
by $a_{\tau}$ and all the left-moving fluxes by $-a_{\tau}$. It is now
easy to see that, due to the periodicity in the $\tau$-direction, if we
multiply together $N_{\tau}$ such column transfer matrices, starting with
$M_{0,2}$, we get the identity matrix: $M_{2N_{\tau}-1} = {\bf 1}$. This
is exactly the situation shown on Fig.~\ref{fig1}, where we chose $N = 2
N_{\tau} - 1 = 5$.

When we introduce a small amount of back scattering, taking $\alpha =
\delta \ll 1$ and $\gamma = - \beta =1 + {\cal O}(\alpha^2)$, the total
transfer matrix $M_{2N_{\tau}-1}$ is still close to the identity matrix. 
However, the back scattering will produce nonzero off-diagonal elements.
The paths contributing to the elements of blocks $m_2$ and $m_3$ of the
matrix $M_{2N_{\tau}-1}$ will have at least one back scattering event on
them, and, therefore, all the matrix elements of $m_2$ and $m_3$ will be
of the order ${\cal O}(\alpha)$. Similarly, the off-diagonal elements in
$m_1$ and $m_4$ will come from the paths having at least two back
scattering events, and will be of the order ${\cal O}(\alpha^2)$. 

When we reinstate the link phases, the matrix elements of the matrix
$M_{2N_{\tau}-1}$ will also acquire some phase factors. These phases are
correlated, because the paths contributing to two different matrix
elements may have links in common. However, the resulting transfer matrix
must be pseudo-unitary due to current conservation and can be factorized
into a product, consisting of a real matrix sandwiched between two
block-diagonal unitary matrices The real factor in this decomposition is
of the form described in the previous paragraph. Thus, the back scattering
is of the same order between any right- and left-moving channel on a
mesoscopic scale $2N_{\tau} a_x$, which is much smaller than the
localization length $\xi$ of Eq.~(\ref{loclength}), because in the limit
we are discussing now $t^2/(1-t^2)=(\beta/\alpha)^2 \gg 1$. This equal
mixing among the channels, happening in the metallic regime $L \ll \xi$,
well before localization sets in, is the essential property of quasi-1D
systems. By contrast, in isotropic 2D systems, like the original
Chalker-Coddington model~\cite{cc}, the back scattering mixes the channels
only locally in the transverse direction, and when the system is at the QH
critical point, the value of the length $L$ at which the channels are all
equally mixed is of the order of the width $W$. For such systems with $L
\gg W$ localization sets in with localization length $\xi \sim W$, so
there is no metallic quasi-1D regime. Such a regime occurs when a
separation of scales $\xi \gg W$ takes place, but only if the
dimensionless 2D conductivity $\sigma_{xx}$ is large, whereas at the QH
critical point $\sigma_{xx}$ takes on a universal value of order 1. For the DN
model the crossover to 1D behaviour occurs at $L \sim L_0 \sim C^{1/2}$, which
is much less than the localization length $\xi \sim C$, so a 1D metallic
regime occurs. 

In quasi-1D wires (see~\cite{smmp} for a review of the scattering approach
to disordered conductors), one starts with the idea that a wire of length
$L$ can be obtained by combining many short (mesoscopic) building blocks
of length $\delta L$. The transfer matrix $M_L$ for the wire is a product
of the transfer matrices $M_{\delta L}$ for the individual blocks.  The
parameters $x_n$, introduced above, are simply related to the eigenvalues
of the transfer matrices. Upon multiplication of the transfer matrices of
the building blocks, the parameters $x_n$ perform a random walk, and their
probability distribution for the total transfer matrix is obtained by
repeated convolution of the distribution of $x_n$ for the building blocks.
Upon taking a limit, in which $M_{\delta L} \to {\bf 1}$ as $\delta L \to
0$, the distribution of the $x_n$ is described by a universal FP equation
with $L$ a continuous variable, this is the DMPK equation. In our system,
the building block is the cylindrical DN of length $2N_{\tau} a_x$.
Therefore, the universal properties of the DN model in the quasi-1D
scaling limit should coincide with those of the DMPK equation, at least
when $t^2/(1-t^2) \gg 1$.

Finally, we want to compare our results with those of Chalker and
Dohmen~\cite{cd}. They introduced an amplitude ratio $A_{\rm CD} \equiv
\xi_{\rm CD}/C$ and found it was equal to $t^2/(1-t^2)$ in agreement with
numerical simulations, while from our Eq.~(\ref{loclength}) we find for
this ratio $A = 2t^2 a_x/(1-t^2) a_{\tau}$. First, we should point out
that Chalker and Dohmen measured their $L$ and $C$ in units of $a_x$ and
$a_{\tau}$, respectively. This takes care of the factor $a_x/a_{\tau}$.
We attribute the remaining factor of~2 difference between $A$ and $A_{\rm
CD}$ to the different conventions used in the definitions of the
localization length~$\xi$. It appears from the equations of Chalker and
Dohmen that they defined their localization length $\xi_{\rm CD}$ through
the decay of the typical conductance in the localized regime, as
$\mean{\ln g} = - L/\xi_{\rm CD}$, while we used the more conventional
definition, such that Eq.~(\ref{logg}) holds. Then our $\xi = 2 \xi_{\rm
CD}$, which explains the factor of~2.

There is one more discrepancy between our results and those of Chalker and
Dohmen. From their expressions for the conductance one finds that the
universal number of Eq.~(\ref{univnum}) is $-2$. This may be explained as
follows. Chalker and Dohmen used rescaled parameters $\nu_n$, introduced
above, and made a conjecture (seemingly based on the 1D model) that the
value of $\nu_1$ is the same in the metallic and localized limits. For the
1D model this conjecture does not hold, as we saw above. If instead of
this conjecture Chalker and Dohmen had assumed that the $\nu_n$'s behave
in the same way as in the 1D model, i.e. that $(\nu_1)_{L \ll \xi} = 2
\mean{\nu_1}_{L \gg \xi}$, as predicted by our mapping, their expressions
would completely agree with Eq.~(\ref{univnum}). We conclude that our
results agree with the numerics of Chalker and Dohmen, and explain why
their heuristic argument works (after factors of~2 are corrected).

\section{Conclusion}
\label{5}

In conclusion, we have considered the directed network (DN) of edge states
on the surface of a cylinder. We claim that, in a certain scaling limit,
the DN is equivalent to the 1D supersymmetric non-linear sigma model and to
the random matrix model used before to describe the transport properties of
quasi-1D wires. Using the known results for this 1D model, we obtain a
description of the conductance properties of the DN model in this scaling
limit. In particular, we give exact expressions for the mean conductance
$\mean{g}$ and the correlation length $\xi$ of the DN model in the scaling
limit for {\it any\/} value of the ratio $C/L$ of the circumference and the
length of the cylinder, Eqs.~(\ref{cond4}, \ref{loclength}), while
expressions for the variance and the distribution of the conductance may be
found in the literature on quasi-1D wires~\cite{mmz,br,mc}. These results
are universal; in particular, they should not depend on the precise
distribution chosen for the disorder (provided its correlations are
short-ranged). They are valid except for short systems with $L \sim L_0
= (C \rho_s / M_0)^{1/2}$ or less, which are expected to behave as 2D
chiral metals. In essence, the crossover from 2D to quasi-1D behaviour is
governed by anisotropic scaling, and hence occurs at $L \sim L_0 \sim
C^{1/2}$, while localization sets in at larger scales, $L \sim \xi \sim C$.

\section{Acknowledgements}

We thank J.~T.~Chalker for helpful correspondence. This research was supported
by NSF grants, Nos. DMR--91--57484 and DMR--96--23181.

\nopagebreak

\clearpage



\end{document}